# Epitaxial Growth of β-Ga$_2$O$_3$ Coated Wide Bandgap Semiconductor Tape for Flexible UV Photodetector

*Xiao Tang, Kuang-Hui Li, Yue Zhao, Yanxin Sui, Huili Liang, Zeng Liu, Che-Hao Liao, Zengxia Mei, Weihua Tang, Xiaohang Li\**

Xiao Tang, Kuang-Hui Li, Che-Hao Liao, Xiaohang Li
Advanced Semiconductor Laboratory, King Abdullah University of Science and Technology (KAUST), Thuwal, 23955-6900, Saudi Arabia
E-mail: xiaohang.li@kaust.edu.sa

Yue Zhao
School of Electronic Information and Electrical Engineering, Shanghai Jiao Tong University, 200240, Shanghai, China

Yanxin Sui, Huili Liang, Zengxia, Mei
Key Laboratory for Renewable Energy, Beijing Key Laboratory for New Energy Materials and Devices, Beijing National Laboratory for Condensed Matter Physics, Institute of Physics, Chinese Academy of Sciences, Beijing 100190, China

Zeng Liu, Weihua Tang
Laboratory of Information Functional Materials and Devices, School of Science and State Key Laboratory of Information Photonics and Optical Communications, Beijing University of Posts and Telecommunications, Beijing 100876, China

The epitaxial growth of technically-important *β*-Ga$_2$O$_3$ semiconductor thin films have not been realized on flexible substrates due to limitations by the high-temperature crystallization conditions and the lattice-matching requirements. In this report, for the first time single crystal *β*-Ga$_2$O$_3$ (-201) thin films is epitaxially grown on the flexible CeO$_2$ (001)-buffered hastelloy tape. The results indicate that CeO$_2$ (001) has a small bi-axial lattice mismatch with *β*-Ga$_2$O$_3$ (-201), thus inducing a simultaneous double-domain epitaxial growth. Flexible photodetectors are fabricated based on the epitaxial *β*-Ga$_2$O$_3$ coated tapes. Measurements show that the obtained photodetectors have a responsivity of 40 mA/W, with an on/off ratio reaching 1000 under 250 nm incident light and 5 V bias voltage. Such photoelectrical performance is already within the mainstream level of the β-

Ga$_2$O$_3$ based photodetectors by using the conventional rigid single crystal substrates; and more importantly remained robust against more than 1000 cycles of bending tests. In addition, the epitaxy technique described in the report also paves the way for the fabrication of a wide range of flexible epitaxial film devices that utilize the materials with lattice parameters similar to β-Ga$_2$O$_3$, including GaN, AlN and SiC.

## 1. Introduction

In the past decade, UV (100-400 nm) photodetectors (PDs) have attracted widespread attention due to their various potential applications such as photolithography, biological identification, communication, and flame detection.[1-10] In general, the configuration of such UV PDs essentially relies on the embedded wide band gap semiconductor layer, which is usually made of compounds such as AlGaN, MgZnO, Ga$_2$O$_3$, and diamond.[3, 8-10] Among these candidate materials, β-Ga$_2$O$_3$ shows several appealing advantages. First of all, β-Ga$_2$O$_3$ has an intrinsic wide band gap of 4.5-4.9 eV, without the need of additional element alloying.[3] Secondly, β-Ga$_2$O$_3$ has excellent thermal and chemical stability, making it promising in harsh environment applications.[4]

By far, the fabrication of β-Ga$_2$O$_3$ thin film based PDs have been reported by numerous research groups and developed into a fairly mature technique. It is known that excellent photodetection characteristics of such PDs require accurate control over the composition, uniformity, and crystallinity of β-Ga$_2$O$_3$ thin films. Up till now, the ever-reported highly crystallized β-Ga$_2$O$_3$ thin films have to be epitaxially grown on quite limited kinds of rigid single crystal substrates with relatively small lattice-mismatch with β-Ga$_2$O$_3$, including native β-Ga$_2$O$_3$ substrate, sapphire (α-Al$_2$O$_3$) (001), MgO (110), and 3C-SiC (001).[11-14] Besides, crystallization of β-Ga$_2$O$_3$ has to be realized by using high-temperature (500~1200 °C) deposition routes, such as molecular beam epitaxy (MBE), pulsed laser deposition (PLD), mist chemical vapor deposition (mist CVD), and metal-organic chemical vapor deposition (MOCVD).[14-19] The harsh condition in these deposition processes sets additional thermal-stability requirement for the growing substrates.

On the other hand, portable and wearable PD devices have been emerging in the recent years, since they can provide people with real-time monitoring of various potential environmental and health hazards regardless of location and time.[20-24] Such applications require the entire micro-electronic system mechanically flexible to accommodate applied bending stress.[25] Obviously, the aforementioned single crystal substrates can hardly meet this requirement. For this purpose, efforts have been made to transfer preformed single crystal β-$Ga_2O_3$ layers to various flexible substrates through the "lift-off-paste" process.[26] However, this technique is not ideal for large-scale production, because of its low area output and low reproducibility. Additionally, several groups have attempted to directly grow $Ga_2O_3$ film on flexible substrates. For instance, Du et.al deposited $Ga_2O_3$ film on flexible polyethylene naphthalene (PEN) polymer substrates.[27] Due to the intrinsic low glass transition temperature of PEN, the deposition was realized via sputtering at room temperature. This made the obtained $Ga_2O_3$ thin films remained in amorphous phase, which could be unstable for long-term device applications.[4, 5, 28] Alternatively, Tang et al used silica glass fiber fabric as the flexible template.[4] The high temperature resistance of silica allows for crystallization of β-$Ga_2O_3$ layer through chemical vapor deposition (CVD) at 900°C. However, since the amorphous silica cannot serve as an epitaxial template, the formed β-$Ga_2O_3$ layer showed no preferred orientation but remained as randomly oriented nanowires.

Overall, till now highly crystalized flexible β-Ga$_2$O$_3$ thin films have not been realized. To achieve this, it becomes imperative to provide a substrate which can simultaneously have excellent bending performance, ideal crystal structure that suitable for epitaxial growth of β-Ga$_2$O$_3$, and robust thermal stability that allows for high temperature reaction.

In this report, for the first time we successfully demonstrated the epitaxial growth of β-Ga$_2$O$_3$ thin films on CeO$_2$ (001)-buffered flexible hastelloy substrate. Here, the hastelloy substrate made of nickel-molybdenum-chromium superalloy with an addition of tungsten, has not only excellent bending performance but also robust thermal stability at a temperature up to 1300 °C.[29] Texture and consequently the epitaxial capability was rendered to the hastelloy substrate through the deposition of multilayers comprising Al$_2$O$_3$/Y$_2$O$_3$/MgO/LaMnO$_3$/CeO$_2$ (001) in consequence by using a combination of sputtering and ion beam assisted deposition (IBAD) techniques. After that, the growth of β-Ga$_2$O$_3$ thin films are realized by PLD on top of the CeO$_2$ layer. The X-ray diffraction (XRD) 2-theta scan shows that the β-Ga$_2$O$_3$ thin film has high phase purity with single out-of-plane orientation. On the other hand, the XRD phi-scan shows a twelvefold symmetry of the grown β-Ga$_2$O$_3$ thin film. It reveals that CeO$_2$ (001) has a bi-axial small-lattice mismatch with the β-Ga$_2$O$_3$ (-201) plane, inducing a simultaneous double-domain growth of β-Ga$_2$O$_3$ thin film. The prepared β-Ga$_2$O$_3$ thin films are fabricated into PDs by depositing micrometer-scaled Ti/Au contacts. The obtained PDs showed a responsivity of 40 mA/W, with an on/off ratio reaching 1000 under 250 nm incident light and 5 V bias voltage. Such photoelectrical performance is already within the mainstream level of the β-Ga$_2$O$_3$ based PDs by using the conventional rigid single crystal substrates; and more importantly remained robust against more than 1000 cycles of bending tests.[3]

The excellent semiconducting and mechanical performances suggest that the epitaxial β-Ga₂O₃ layer coated tape is applicable for not only flexible PDs but also other deformable devices such as flexible Ga₂O₃ thin film transistor (TFT) and wearable Ga₂O₃ gas sensors. Moreover, the epitaxy technique described in this report paves the way for the fabrication of a wide range of flexible epitaxial film devices that utilize the materials with lattice parameters similar to β-Ga₂O₃, including GaN, AlN and SiC.

## 2. Results and Discussion

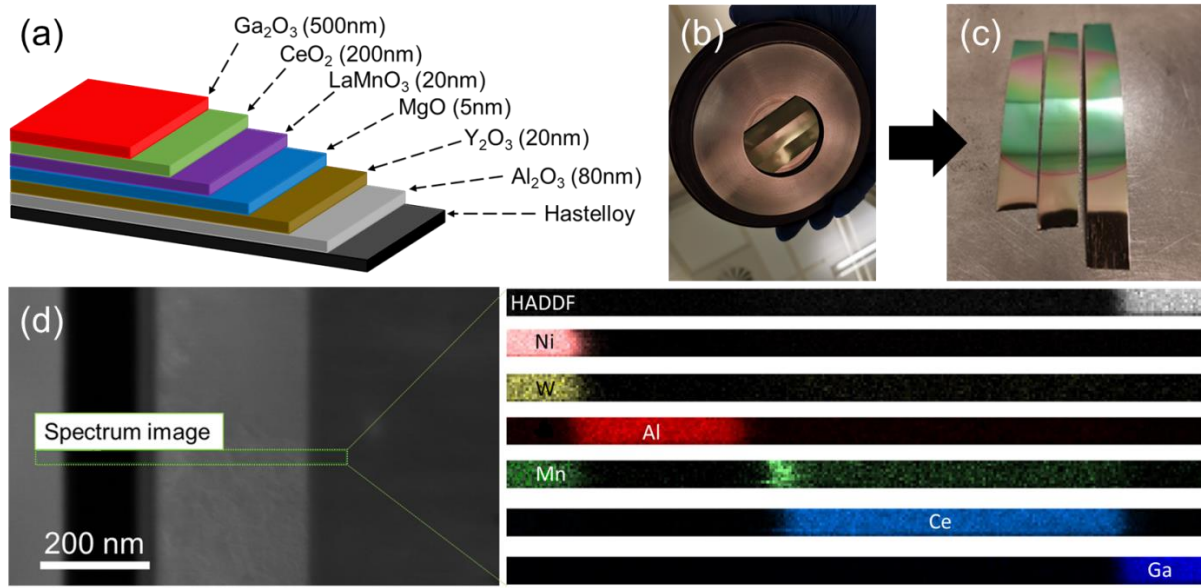

**Figure 1.** (a) Cross-sectional schematics showing the layer configuration of the β-Ga₂O₃ coated flexible tape; (b) photograph of a batch of the CeO₂ buffered hastelloy tapes mounted on the PLD sample holder with a two-inch deposition opening; (c) photograph of the same tapes after the β-Ga₂O₃ deposition; (d) Cross-sectional HRTEM, HADDF and EDX images of the β-Ga₂O₃ coated tape.

First, the thermal stability of the hastelloy tape and the epitaxial ability of CeO₂ (00l) thin film for growing β-Ga₂O₃ (-201) layer are tested as shown in Figure S1 and S2 in the Supporting

Information, respectively. By applying the IBAD technique, the hastelloy tape is functionalized with textured $CeO_2$ layer that of (00l) preferred orientation. Then β-$Ga_2O_3$ is deposited on top of the $CeO_2$ layer. **Figure 1** (a) illustrates the structural configuration of the β-$Ga_2O_3$ coated tape. In general, the stacked structure is built-up based on a C-276 hastelloy tape with 10 mm in width and 50 μm in thickness. After an electropolishing procedure, the tape is first coated with amorphous $Al_2O_3$ (80 nm) and $Y_2O_3$ (20 nm) layers successively by using reactive RF sputtering. Here, the $Al_2O_3$ layer serves as a barrier layer to minimize the diffusion of the metallic elements during the following high-temperature treatments; while the $Y_2O_3$ layer serves as a seed layer for the following depositions of the crystallized epitaxial layers. As the next step, highly crystalized MgO layer (5 nm) with (00l) orientation is realized by the IBAD technique. After that, $LaMnO_3$ (20 nm) and $CeO_2$ (200 nm) layers are successively deposited by magnetron sputtering. Thanks to the epitaxy that provided by the crystallized MgO layer, the deposited $LaMnO_3$ and $CeO_2$ layers naturally obtain a preferred orientation along (00l). Till this step, the deposition of the stacked buffer-layer structure is accomplished. Then the β-$Ga_2O_3$ thin film (500 nm) is deposited on top of the $CeO_2$ layer via PLD. It is worth mentioning that systematic thickness optimization of the β-$Ga_2O_3$ thin film is performed and the results are shown in Figure S3 in the Supporting Information. Figure 1 (b) and (c) show a batch of the buffered hastelloy tapes before β-$Ga_2O_3$ deposition that mounted on a PLD sample holder with a 2-inch opening and the same tapes after β-$Ga_2O_3$ deposition (reflected from the greenish area), respectively. To investigate the element distribution along the cross-section of the β-$Ga_2O_3$ coated semiconductor tape, the high-resolution transmission electron microscopy (HRTEM), the high-angle annular dark-field (HADDF) imaging together with the energy dispersive X-Ray spectroscopy (EDX) are carried out with the results shown in Figure 1 (d). The results indicate distinct interfaces between each layer throughout the stacked structure. Specifically, no diffusion of cerium from the CeO$_2$ layer to the β-Ga$_2$O$_3$ layer is observed. Therefore the material purity of the β-Ga$_2$O$_3$ layer should have been ensured.

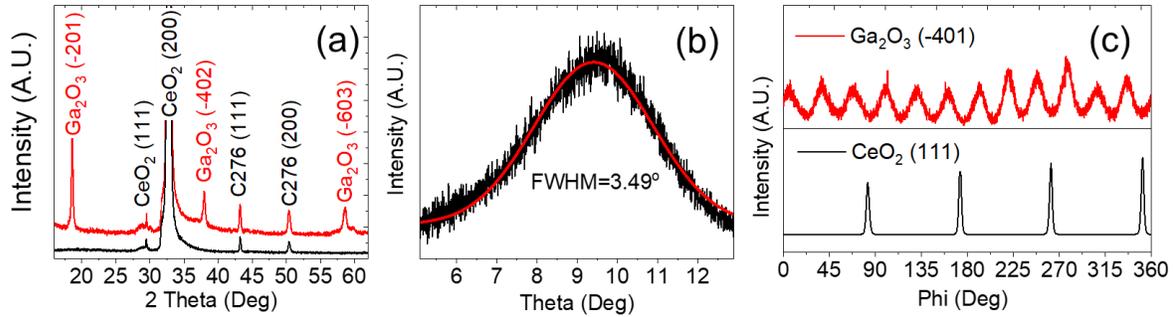

**Figure 2.** (a) Out-of-plane 2 theta XRD patterns of the β-Ga$_2$O$_3$ coated tape (red) and the CeO$_2$ buffered hastelloy tape (black); (b) rocking curve of β-Ga$_2$O$_3$ (-201) reflection of the β-Ga$_2$O$_3$ coated tape, where the red plot represents the Gaussian fitting of the measured data points (black); (c) phi-scan of β-Ga$_2$O$_3$ (-401) plane (red) and CeO$_2$ (111) plane (black) of the β-Ga$_2$O$_3$ coated tape. It is noted that the tape sample was fixed on the XRD sample holder without position change during the phi-scans of Ga$_2$O$_3$ (-401) plane and CeO$_2$ (111) plane.

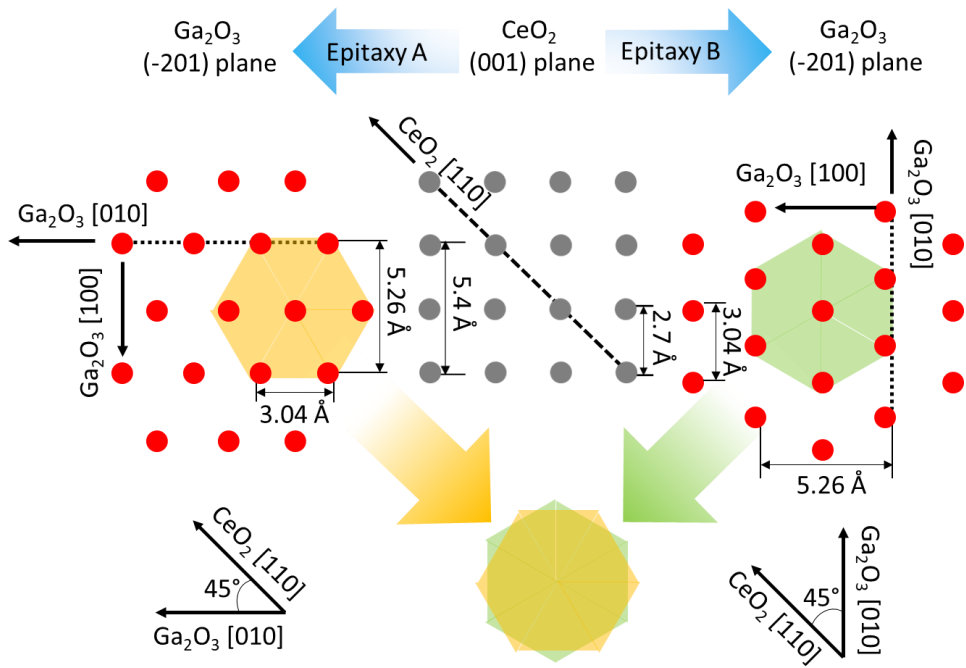

**Figure 3.** Proposed in-plane epitaxial relationship between CeO$_2$ (001) and Ga$_2$O$_3$ (-201). The red dots and the gray dots represent the oxygen atom distribution in CeO$_2$ (001) and β-Ga$_2$O$_3$ (-201) planes, respectively. The orange and green areas highlighted in the two domains represent the smallest sixfold symmetric hexagon units in the β-Ga$_2$O$_3$ (-201) plane for epitaxy A and epitaxy B, respectively.

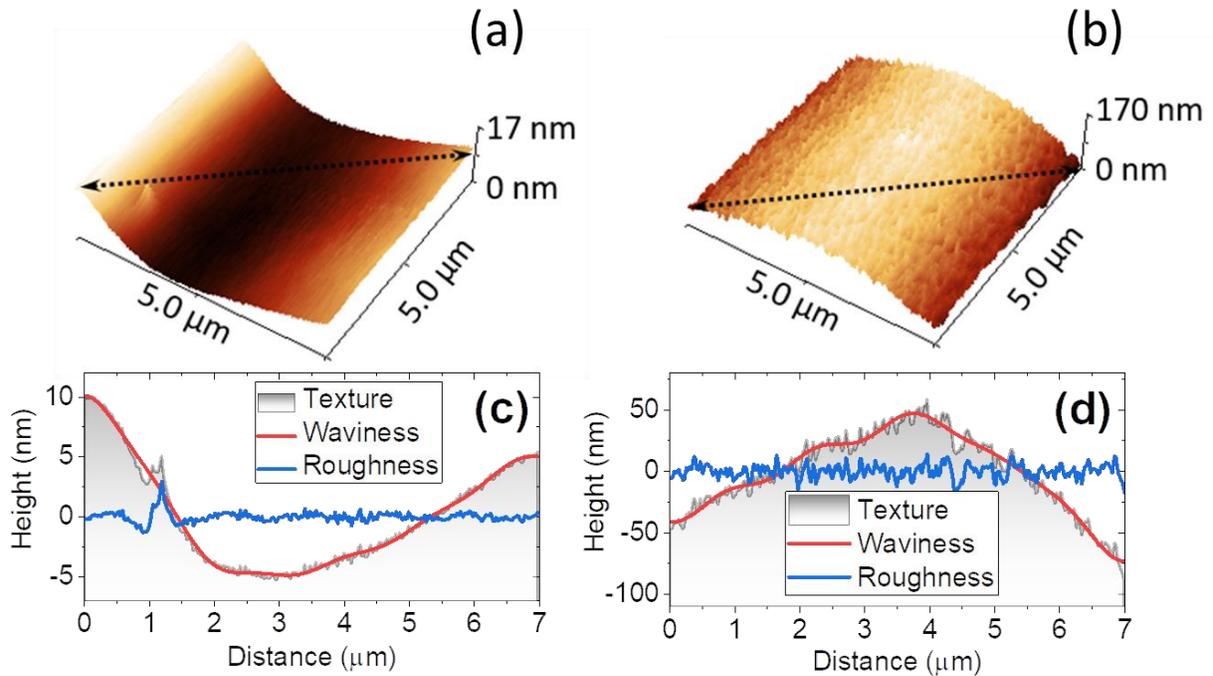

**Figure 4** 3D AFM images (a, b) and height profiles along the black dashed lines (c, d) of the buffered tape and the β-Ga$_2$O$_3$ coated flexible tape, including texture, waviness and roughness.

**Figure 2** (a) shows the XRD 2 theta patterns of the β-Ga$_2$O$_3$ coated tape and the buffered tape. For the buffered tape, the strong (002) peak and the weak (111) peak of CeO$_2$ are observed in the pattern. Besides, another two peaks at higher diffraction angles are corresponding to (002) and (111) diffractions of hastelloy C276. When coated with β-Ga$_2$O$_3$, the pattern is characterized by the appearance of the series of strong diffraction peaks from $\{\bar{2}01\}$ planes of β-Ga$_2$O$_3$, without any other diffraction peaks (e.g. β-Ga$_2$O$_3$ (400) at approx. 30°). The result indicates successful epitaxial growth of single (-201) oriented β-Ga$_2$O$_3$ thin films on the (001) oriented CeO$_2$ buffer layer. In addition, the same 2 theta measurements are performed on the whole deposition area that shown in Figure 1 (c). The results are represented in Figure S4 in Supporting Information and demonstrate high uniformity of the crystallization quality of the samples. Rocking curve is obtained around the (-201) plane of β-Ga$_2$O$_3$ as shown in Figure 2 (b) and gives a full width half maximum (FWHM) value of 3.49°. Indeed, the value is higher compared to those of the β-Ga$_2$O$_3$ thin films deposited on the conventional sapphire single crystal substrates by PLD.[30, 31] This can result from two reasons. First, the beneath CeO$_2$ layer that serves as the epitaxial growing template is not technically a single crystal, but only a highly crystallized thin film with a preferred orientation. This means that the crystal quality, consequently the epitaxy ability of the CeO$_2$ (001) layer is not comparable to those of the sapphire substrates. Secondly, since the thin films are based on the flexible hastelloy substrate, bending is technically inevitable despite pasting on a rigid and flat holder during the XRD measurement. Indeed, such bending is invisible to naked eye, however it can be well detected in atomic force microscopy (AFM), which would be described in **Figure 4**. This can undoubtedly result in a certain de-planarization of the (-201) plane of the β-Ga$_2$O$_3$ thin film, therefore contributing to the larger FWHM value of the rocking curve. To investigate the in-plane epitaxial relationship between the CeO$_2$ (001) layer and the β-Ga$_2$O$_3$ (-201) layer, the β-Ga$_2$O$_3$ coated tape is also subjected to XRD phi-scan measurements, as shown in Figure 2 (c). The phi-scan of CeO$_2$ (111) manifests four peaks, representing a typical fourfold symmetry along the [001] axis of its cubic structure. The β-Ga$_2$O$_3$ film shows 12 reflection peaks with a separation of 30° between each of them. It is known that for single crystal β-Ga$_2$O$_3$ (-201) thin films that epitaxially grown on the sapphire substrates, six peaks evenly separated by 60° are often observed because of the sixfold symmetry along the [102] axis of the crystal structure.[3] Hence, the 12 peaks observed in the sample suggest that the β-Ga$_2$O$_3$ film has two domains; and the β-Ga$_2$O$_3$ (-401) peaks come from such two domains appear alternately through-out the 0-360° scan. It therefore indicates that the two domains are laid perpendicularly in the CeO$_2$ (001) plane, with a rotation angle of 90° between them. In addition, it is observed that all the four CeO$_2$ peaks locate exactly at the midpoint between the two adjacent β-Ga$_2$O$_3$ (-401) peaks. This means that at the β-Ga$_2$O$_3$/CeO$_2$ interface, the intersection between the CeO$_2$ (111) and the CeO$_2$ (001) planes always forms 45° angle with that between the β-Ga$_2$O$_3$ (-401) and the β-Ga$_2$O$_3$ (-201) planes.

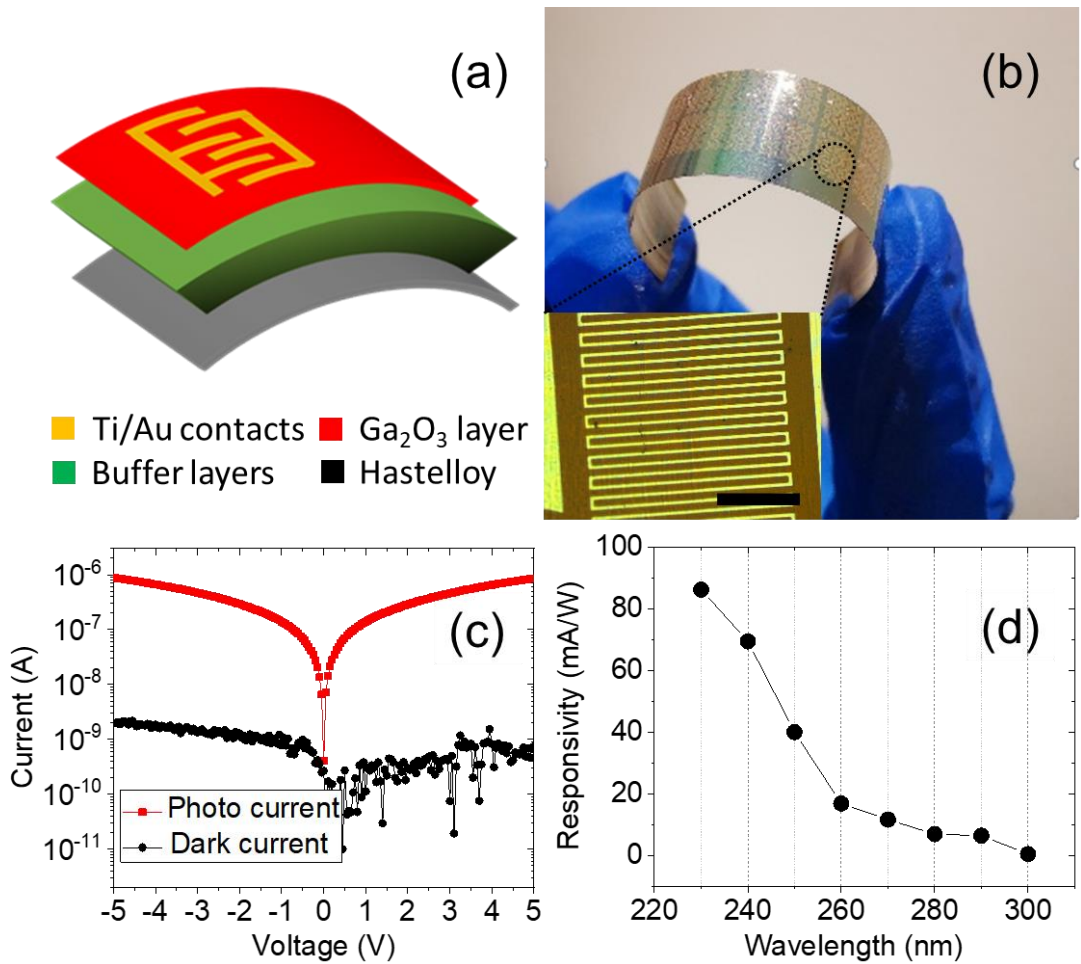

**Figure 5.** (a) Simplified schematic illustration and (b) photograph of the flexible PD based on the β-Ga₂O₃ coated semiconductor tape, and the inset shows a microscopic photograph of the Ti/Au electrode array with a scale bar equals to 100 μm. (c) I-V characteristics of the sample tested in dark and under 254 nm illumination. (d) Wavelength-dependent responsivity of the sample.

Based on the understanding above, the epitaxial relationship between CeO₂ (001) and β-Ga₂O₃ (-201) planes is illustrated in **Figure 3**. The red dots and the gray dots represent the oxygen atom distribution in CeO₂ (001) and β-Ga₂O₃ (-201) planes, respectively. It is known that the intersection between the CeO₂ (111) and the CeO₂ (001) planes coincides with the CeO₂ [110] direction (black dashed line); and the intersections between the β-Ga₂O₃ (-401) and the β-Ga₂O₃ (-201) planes coincides with the β-Ga₂O₃ [010] direction (black dashed line). Therefore, relating to the phi-scan result, the CeO₂ [110] and the Ga₂O₃ [010] directions form 45° angle either clockwise or counterclockwise, resulting in two perpendicularly laid domains, as shown in the left side (Epitaxy A) and the right side (Epitaxy B), respectively. And the orange and green areas highlighted in the two domains represent the smallest sixfold symmetric hexagon units in the β-Ga₂O₃ (-201) plane for Epitaxy A and Epitaxy B, respectively. By superimposing them together, a twelvefold symmetric geometry is observed, which is actually reflected from the phi-scan result. To better understand the mechanism for the epitaxial growth of the two perpendicularly laid β-Ga₂O₃ domains, we carefully looked inside the distance between the neighboring oxygen atoms in the planes of β-Ga₂O₃ (-201) and CeO₂ (001). According to the calculation described in the previous reports, the distances between the neighboring oxygen atoms alone CeO₂ [100] direction is 2.7 Å; while those along the β-Ga₂O₃ [100] and [010] directions are 3.04 and 5.26 Å, respectively.[32-34] The atom distance along the β-Ga₂O₃ [100] direction is approximately corresponding to that along the CeO₂ [100], giving a lattice mismatch of 11%; while the atom distance along the β-Ga₂O₃ [010] is approximately corresponding to two times of that along the CeO$_2$ [100] (5.4 Å), giving a lattice mismatch of only 2.6%. Here it is noted that for the conventional epitaxial growth of β-Ga$_2$O$_3$ thin films on sapphire substrate, the lattice mismatch between β-Ga$_2$O$_3$ [010] and Al$_2$O$_3$ [11$\bar{2}$0] reaches at 10.7%.[3] It therefore means that the lattice mismatch between either β-Ga$_2$O$_3$ [100] or [010] and CeO$_2$ [100] are all small enough for epitaxial growth. As a result, two perpendicularly laid β-Ga$_2$O$_3$ domains form simultaneously on the CeO$_2$ (001) surface.

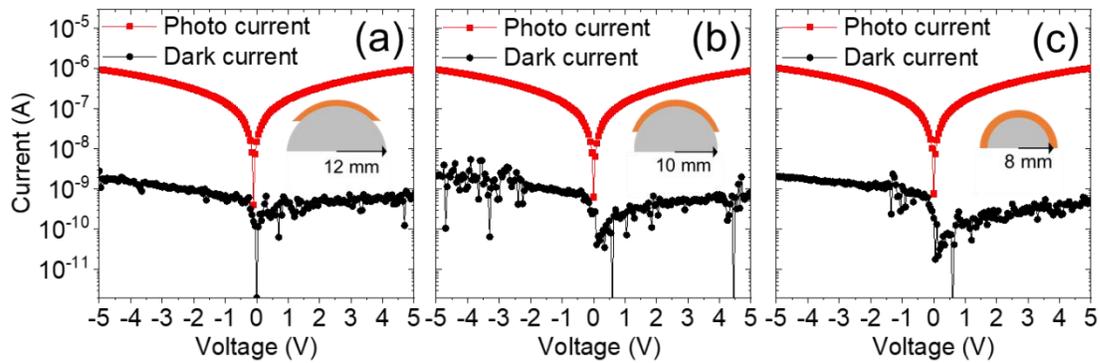

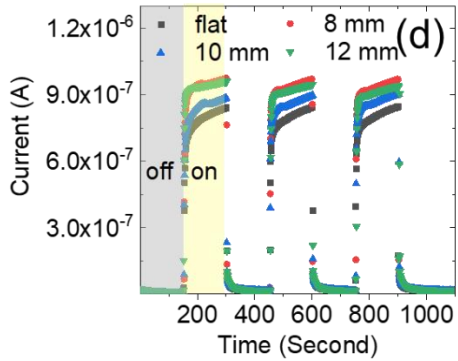
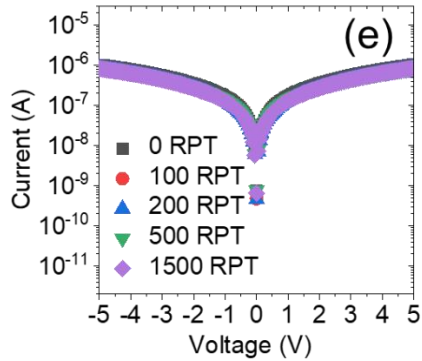

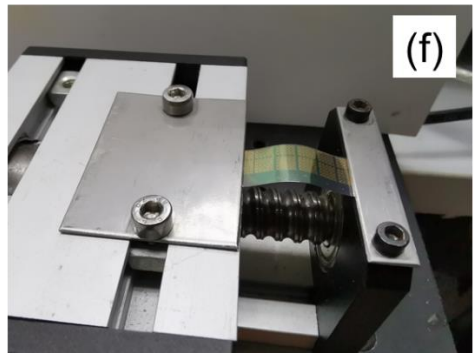
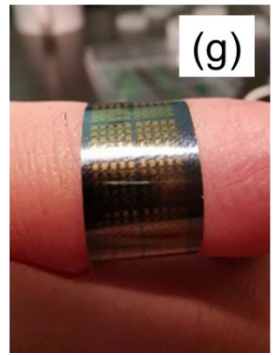

**Figure 6.** (a-c) I–V characteristics of the flexible PD sample under different bending conditions. (d) Time-dependent photoresponse curves measured under 254 nm illumination that turned on/off periodically at 120 s intervals. (e) I-V characteristics accompanied with an inset showing the zoom-in view at 4−5 V of the flexible PD sample under 254 nm illumination after 0, 100, 200, 500 and 1500 times repeated (RPT) bending cycles (r = 10 mm). (f) Photograph of the machine for the bending fatigue test. (g) Photograph of a PD finger ring made by rolling an 8 cm long sample with a bending radii of 10 mm.

Figure 4 (a, b) show representative three-dimensional (3D) AFM images of the buffered tape and the β-Ga$_2$O$_3$ coated tape. The corresponding height profiles measured along the black dashed lines are represented in Figure 4 (c, d). The same as the XRD measurements, the tape samples are also pasted onto a rigid substrate by using double-side tape glue for the AFM scans. Nevertheless, bending of the tapes is evident, since the height change (as quantitatively indicated by geometry of the 3D AFM images and the texture values in the height profiles) in the two samples are both majorly contributed from the waviness rather than the intrinsic roughness of the samples. First, the results strongly support our explanation given for the rocking curve measurement. That is, the relative high FWHM value can be partially due to the bending of the sample. Second, after reduction of the waviness as the baseline, the CeO$_2$ buffered tape and the β-Ga$_2$O$_3$ coated tape shows RMS roughness of 0.4 and 5.5 nm, respectively. Obviously, the roughness significantly increased after β-Ga$_2$O$_3$ deposition; however the value is still fairly low so that the β-Ga$_2$O$_3$ semiconductor tapes are capable of being fabricated into various photoelectronic devices.

Subsequently, the device fabrication is carried out. Coplanar interdigital Ti (50nm)/Au (195nm) electrodes with a spacing of 10 μm were deposited on the β-Ga$_2$O$_3$ coated flexible tape to form metal-semiconductor-metal (MSM)-structured PDs. A simplified configuration and a photograph of the bent flexible PD is shown in **Figure 5** (a) and (b), respectively. Thanks to the low surface roughness of the β-Ga$_2$O$_3$ layer, the deposited Ti/Au electrode array possessed intact structure with smooth surface and clear edges as shown in the microscopic photograph (the inset of Figure 5 (b)). The current–voltage (I-V) characterstics of the PD are shown in Figure 5 (c). The off-state dark current of the device is approximately 1 nA at 5 V. When illuminated by 254 nm UV light with power density of 20 mW/cm$^2$, the on-state photocurrent is increased by three orders of magnitude, reaching approximately 1 μA. The wavelength dependent photoresponse is measured on the PD sample by tuning the illuminating light wavelength in the range of 230-300 nm at a fixed bias voltage of 5 V. Based on the measured results, the photoresponsivity of the PD sample is calculated by using Equation 1:

$$R = \frac{I_{photo}}{D \times S} \qquad (1)$$

where $R$, $I_{photo}$, $D$, and $S$ are the photoresponsivity, photocurrent, illuminating power density and the exposure area, respectively. The calculated photoresponsivity as a function of illuminating wavelength is shown in Figure 5 (d). The photoresponsivity of the PD sample is 0.4 mA/W at 300 nm and then sees a dramatic increase starting from 260 nm, eventually reaching 86 mA/W at 230 nm. The result corresponds well to the theoretical band gap of β-Ga$_2$O$_3$.[3] However, it is noted that the band gap of the deposited β-Ga$_2$O$_3$ layer in this study cannot be simply determined by ultraviolet–visible (UV-Vis) spectroscopy since the hastelloy substrate is not transparent. In general, the PDs fabricated based on the β-Ga$_2$O$_3$ coated hastelloy tape manifests good photoelectrical property in terms of the on/off ratio ($I_{photo}/I_{dark}$ =1000) and the wavelength selectivity in UV region ($R_{\lambda=230\ nm}/R_{\lambda=300\ nm}$ = 215). In comparison with the reports elsewhere, the key device characteristics shown here are within the mainstream level of the β-Ga$_2$O$_3$ PDs by using the conventional rigid single crystal substrates.[35-39]

The flexibility of the PDs on the flexible tape is tested by measuring its photoelectrical properties under different bending conditions, as shown in **Figure 6** (a-c). Again, the test is performed under 254 nm UV light with a power density of 20 mW/cm$^2$. Different bending conditions are achieved by pasting the flexible PD sample on the semicircular cylinder molds with different radii of 8, 10, and 12 mm depicted in the insets of Figure 6 (a-c). Particularly, an 8 cm long PD tape sample can be easily rolled into a "PD ring" on the finger of yours truly when applying a bending radii of 10 mm, as shown in Figure 6 (f). Thus the bending test in this radii range (8-12 mm) is fairly meaningful to the fabrication of potential wearable and deformable devices. As evident from Figure 5 (a-c), the dark current and the photocurrent of the PD sample remained at 1 nA and 1 µA, respectively under different bending conditions, which are the same to those obtained in the flat state.

To test response time and stability of the PD sample, repetitive measurements under the afornentioned bending conditions with an on/off interval of 120 s are then performed as shown in Figure 6 (d). The results show a visible fluctuation of around ±20 % in the on-state current under the different bending conditions. Such small changes were also found in the flexible PDs that using other semiconductor materials such as ZnO, $CH_3NH_3PbI_3$, and CdS, however the reason is unknown.[40-42] Here, we believe that it is due to the illumination power change induced by the vertical displacement of the flexible sample during the bending measurements.

At last, to test mechanical robustness of the PD sample, the fatigue test is carried out using a home-made machine as shown in Figure 6 (f). First of all, the sample is fully straightened by the two arms. Then the movable arm starts automatically moving back and forth to repeatedly bend the sample. The corresponding bending radii can be estimated based on the displacement of the movable arm. After different bending cycles, the I-V characteristics of the sample is again measured under the aforementioned illuminating conditions. Figure 6 (e) shows the I-V characteristics obtained after 0, 100, 200, 500 and 1500 bending cycles with radii = 10 mm. Evidently, no visible degradation can be observed even after 1500 bending cycles, since the I-V curves are almost overlapped to each other. Taken together, the results indicate that the epitaxially grown β-$Ga_2O_3$ flexible PDs have not only good photoelectrical performance but also robust flexibility, which making them promising for the applications in various wearable devices.

## 3. Conclusion

In summary, we successfully demonstrated the epitaxial growth of β-$Ga_2O_3$ semiconductor thin film on $CeO_2$ (001)- buffered flexible hastelloy tape. The XRD results indicate a bi-axial low-lattice mismatch between $CeO_2$ (001) plane and β-$Ga_2O_3$ (-201) plane. Consequently, two perpendicularly laid $Ga_2O_3$ domains grows simultaneously on the $CeO_2$ (001) plane. The β-$Ga_2O_3$ coated tapes were fabricated into flexible PD by depositing Ti/Au contacts. The PDs have both good photonelectrical performance and robust flexibility, therefore making them promising for the applications in various UV sensitive wearable devices. In addition, we believe that the epitaxial technique described here can be also extended to the fabrication of other technically important devices that based on flexible highly crystallized β-$Ga_2O_3$ thin films such as flexible thin film transistor, power rectifiers and gas sensors. Moreover, the epitaxy between cubic $CeO_2$ and monoclinic β-$Ga_2O_3$ suggests that the application of the $CeO_2$ (001) coated hastelloy tape can be also extended to the fabrication of a wide range of flexible epitaxial film devices that utilize the materials with lattice parameters similar to β-$Ga_2O_3$, such as GaN, AlN and SiC.

## 4. Experimental Section

*Deposition of stacked buffer layers*: The reactive sputtering together with IBAD technique were used to deposit hastelloy/$Al_2O_3$/$Y_2O_3$/MgO/$LaMnO_3$/$CeO_2$ multilayers. The details involved in the above procedures can be found in our previous works.[43, 44]

*Deposition of β-$Ga_2O_3$ layer*: The β-$Ga_2O_3$ thin films were synthesized via PLD. During the deposition process, the $CeO_2$-buffered tapes were cut into 8 cm in length then fixed by two hollow plints. The working distance between the β-$Ga_2O_3$ target and the hollow plints was set at 9.5 cm. Before the deposition starts, the temperature of sample holder and the oxygen partial pressure reached at 640 °C and 5 mTorr, respectively. During the deposition, a KrF excimer laser was operating at a frequency of 5 Hz with an energy per pulse of 300 mJ. The deposition process was finished after 5000-30000 pulses.

*Cross-section characterization*: The HRTEM was applied to investigate the specimen by operating an FEI Titan ST system at an acceleration voltage of 300 kV; and the images were processed by Gatan DitigtalMircrogaph. The TEM specimens were prepared by using an FEI Helios G4 dual-beam focused ion beam scanning electron microscope (DBFIB-SEM) system equipped with an Omni probe and a gallium ion source.

*XRD measurements:* The crystal structure properties of all the described samples were measured by a Bruker D8 ADVANCE X-ray diffractometer using Cu Kα ($\lambda = 1.54$ Å) radiation.

*Surface morphology measurements:* Surface morphologies of all the samples were measured by using a Bruker Dimension Icon SPM AFM.

*PD device fabrication:* The flexible PDs were constructed following the widely-used photolithography standard operation procedures.

*Photoelectrical measurements:* All the I–V curves were measured under 254 nm DUV light. The responsivity measurements were carried out by using a high power mercury-Xenon light sources from Newport as the light source. A Keithley 6487 picoammeter was used as the power supply.

**Supporting Information**

**Acknowledgements**

The authors would like to acknowledge the support of KAUST Baseline BAS/1/1664-01-01, KAUST Competitive Research Grant URF/1/3437-01-01, URF/1/3771-01-01, GCC Research Council REP/1/3189-01-01, and National Natural Science Foundation of China (61874139). The authors would also like to acknowledge the support from Ulrich Buttner in nanofabrication core lab, KAUST and Laurentiu Braic in thin film deposition core lab, KAUST for device fabrication and PLD maintenance, respectively.

**Epitaxial Growth of β-Ga₂O₃ Coated Wide Bandgap Semiconductor Tape for Flexible UV Photodetector**

*Xiao Tang, Kuang-Hui Li, Yue Zhao, Yanxin Sui, Huili Liang, Zeng Liu, Che-Hao Liao, Zengxia Mei, Weihua Tang, Xiaohang Li\**

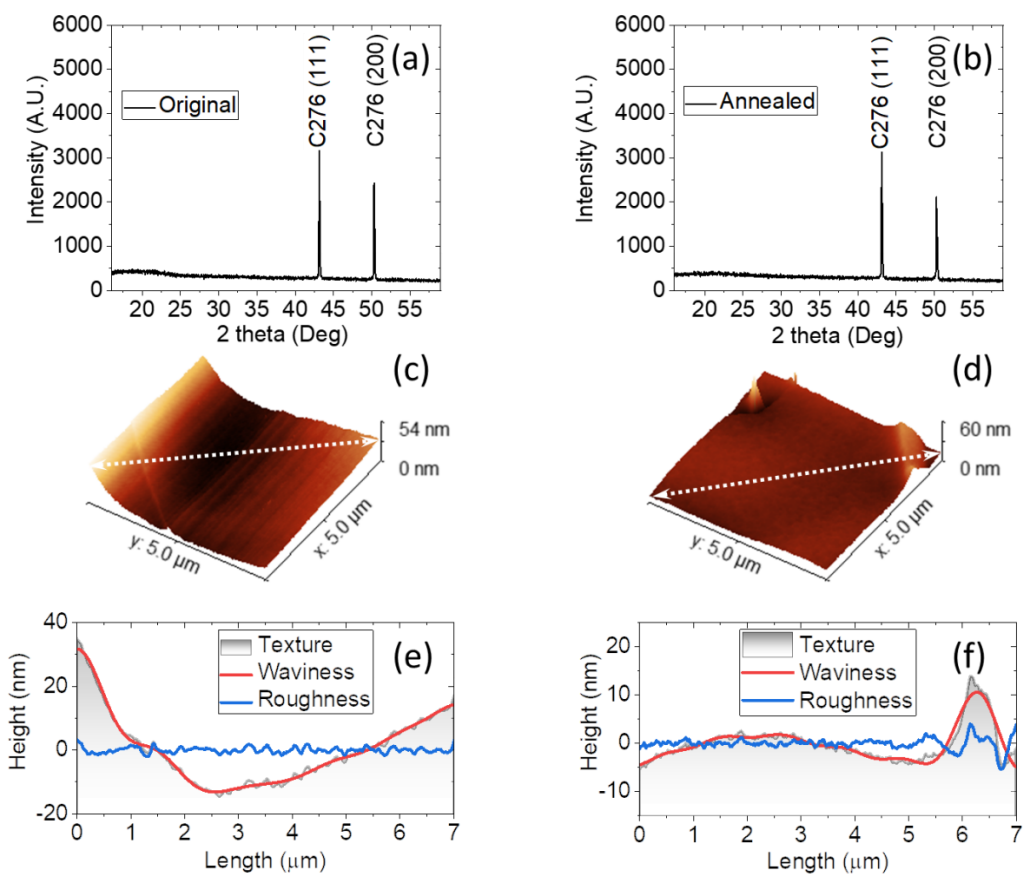

**Figure S1.** (a, b) Out-of-plane 2 theta XRD patterns and (c, d) 3D AFM images of the hastelloy tape before and after annealing at 800 °C for 3 hours. (e, f) Height profiles along the white dashed lines in (c, f), including texture, waviness and roughness. Figure (e) and (f) give root-mean-square (RMS) roughnesses of 0.9 and 1.2 nm, respectively. The results indicate that the hastelloy tape is robust against high temperature conditions at least up to 800 °C, since the high temperature

treatment did not result in considerable change in either XRD pattern or surface roughness. It means the high thermal stability mkaes the hastelloy tape capable of serving as the substrate for the deposition of β-Ga$_2$O$_3$ thin film at a high temperature (640 °C).

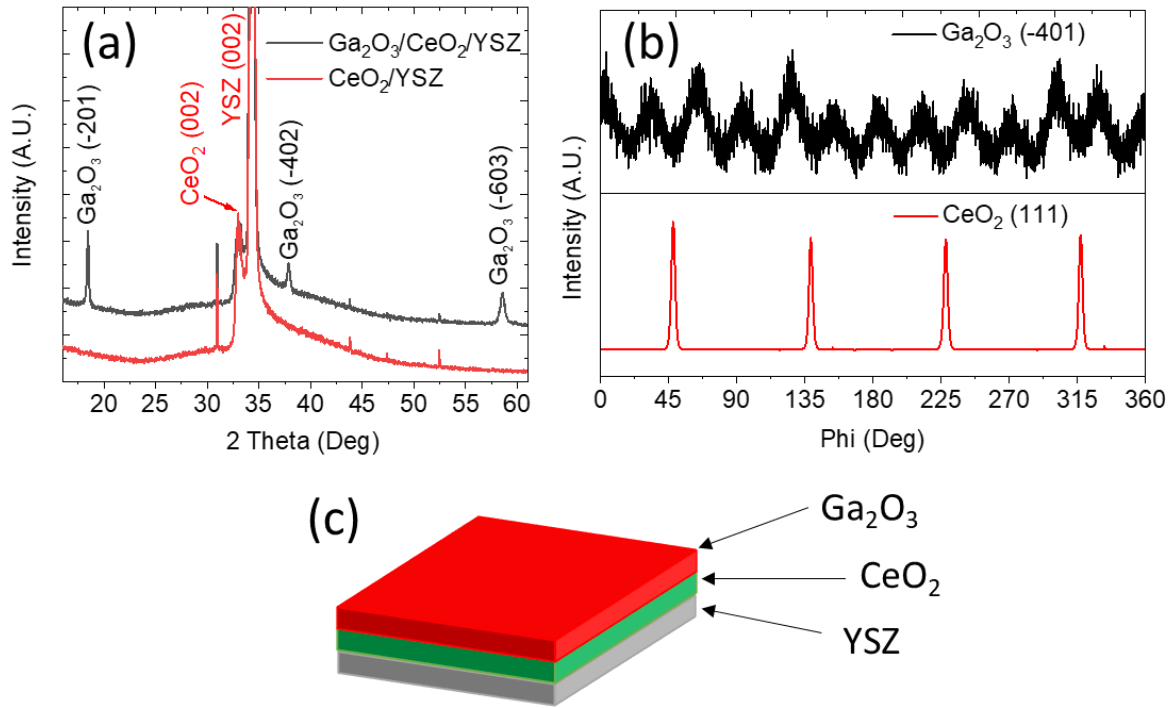

**Figure S2.** (a) Out-of-plane 2 theta XRD patterns of the CeO$_2$/Yttria-stabilized zirconia (YSZ) commercial single crystal substrate (perchased from MTI Technology Corporation) with and without β-Ga$_2$O$_3$ coating; (b) phi-scan of β-Ga$_2$O$_3$ (-401) plane and CeO$_2$ (111) plane of the β-Ga$_2$O$_3$ coated tape. The tape sample was fixed on the XRD sample holder without position change during the phi-scan measurement of β-Ga$_2$O$_3$ (-401) plane and CeO$_2$ (111) plane. (c) Structural configuration of the Ga$_2$O$_3$/CeO$_2$/YSZ buffered structure. The results demonstrate the epitaxial ability of CeO$_2$ (001) surface for the growth of (-201) oriented β-Ga$_2$O$_3$ thin film.

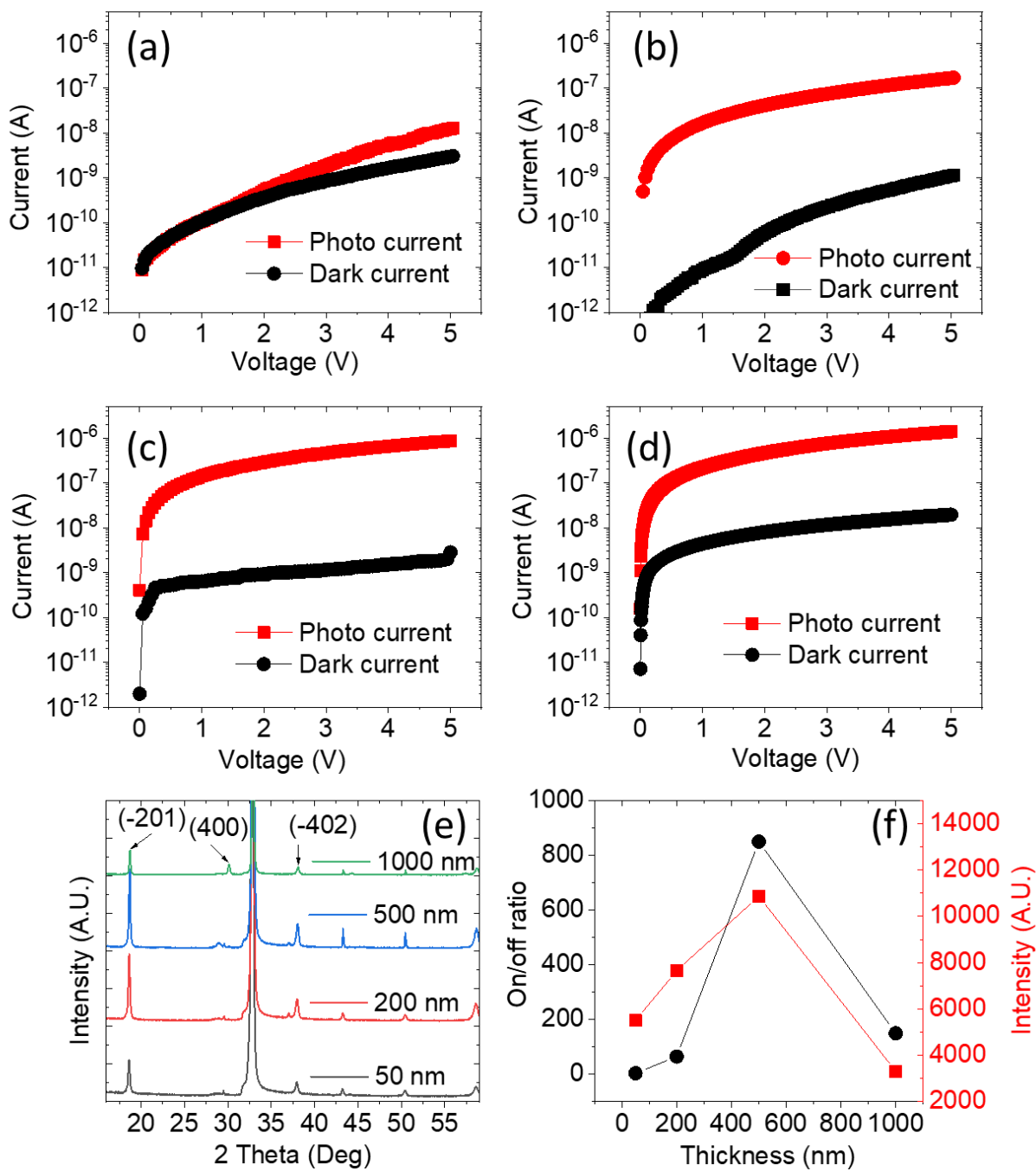

**Figure S3.** (a-d) I-V curves ((a) 50 nm, (b) 200 nm, (c) 500 nm, and (d) 1000 nm) and (e) Out-of-plane theta–2 theta XRD patterns measured on the samples with different β-$Ga_2O_3$ thickness. (f) Evolution of the on/off ratio and the intensity of β-$Ga_2O_3$ (-201) diffraction peak with film thickness. The results indicate that the sample coated with 500 nm $Ga_2O_3$ film has an optimized

crystal structure in term of the intensity of β-Ga$_2$O$_3$ (-201) diffraction peak and consequently a superior photoelectrical performance in term of the on/off ratio. Therefore, the 500 nm sample was chosen for the systematical analysis in our research. In addition, the result reveals that high epitaxial quality is essential to the ideal photoelectrical performance of β-Ga$_2$O$_3$ thin films.

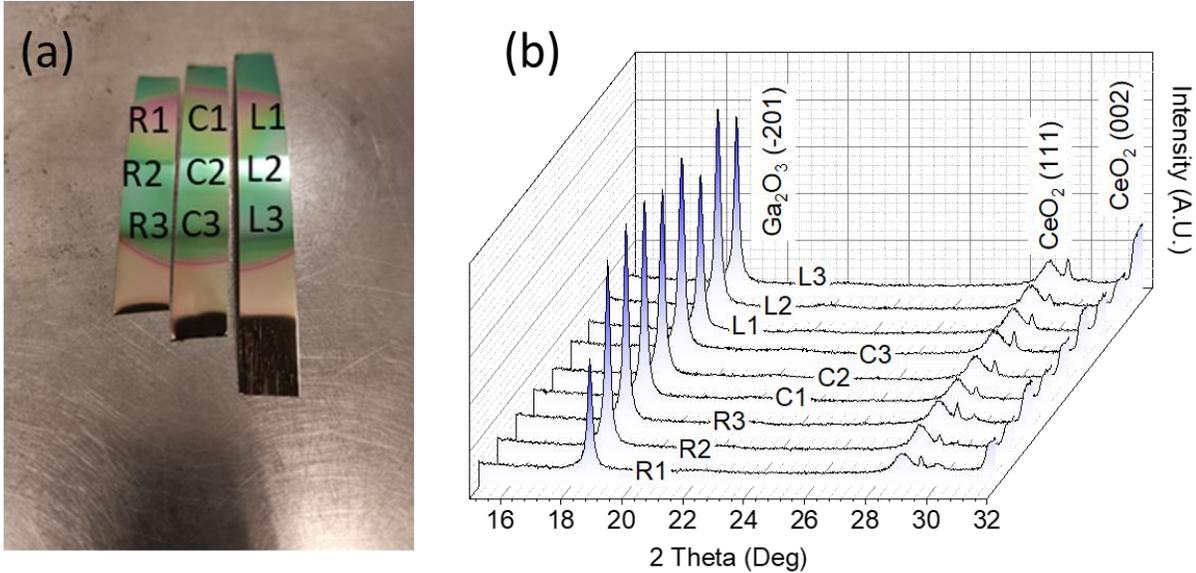

**Figure S4.** (a) Photograph of a batch of samples including three tapes that took out from the PLD chamber after β-Ga$_2$O$_3$ deposition, which is the same as Figure 1(c) except with the position indicators. (b) Out-of-plane 2 theta XRD patterns measured at the various positions marked in (a).